\documentclass[conference]{IEEEtran}
\IEEEoverridecommandlockouts
\usepackage{cite}
\usepackage{amsmath,amssymb,amsfonts}
\usepackage{algorithmic}
\usepackage{graphicx}
\usepackage{textcomp}
\usepackage{xcolor}
\usepackage{tikz}
\usetikzlibrary{arrows, arrows.meta, calc}
\usepackage{todonotes}
\usepackage{hyperref}
\def\BibTeX{{\rm B\kern-.05em{\sc i\kern-.025em b}\kern-.08em
    T\kern-.1667em\lower.7ex\hbox{E}\kern-.125emX}}
\begin{document}

\title{From CCS-Planning to Testautomation:\\The Digital Testfield of Deutsche Bahn in Scheibenberg - A Case Study\\
}

\author{
\IEEEauthorblockN{Arne Boockmeyer\IEEEauthorrefmark{1}, Dirk Friedenberger\IEEEauthorrefmark{2}, Lukas Pirl\IEEEauthorrefmark{1}, Robert Schmid\IEEEauthorrefmark{1} and Andreas Polze\IEEEauthorrefmark{1}}
\IEEEauthorblockA{\textit{Professorship for Operating Systems and Middleware} \\
\textit{Hasso Plattner Institute, University of Potsdam}\\
Potsdam, Germany \\
Email: \IEEEauthorrefmark{1}{\{firstname.lastname\}}@hpi.uni-potsdam.de, \IEEEauthorrefmark{2}{dirk.friedenberger@guest.hpi.de}}
\\
\IEEEauthorblockN{Heiko Herholz\IEEEauthorrefmark{3}}
\IEEEauthorblockA{\textit{Eisenbahn-Betriebs- und Experimentierfeld Berlin (EBuEf)} \\
\textit{Fachgebiet Bahnbetrieb und Infrastruktur, Technische Universität Berlin}\\
Berlin, Germany \\
Email: \IEEEauthorrefmark{3}{heiko.herholz@tu-berlin.de}}
\\
\IEEEauthorblockN{Gisela Freiin von Arnim\IEEEauthorrefmark{4}, Pedro Lehmann Ibáñez\IEEEauthorrefmark{5}, Torsten Friedrich\IEEEauthorrefmark{6}, Christoph Klaus\IEEEauthorrefmark{6} and Christian Wilhelmi\IEEEauthorrefmark{6}}
\IEEEauthorblockA{\textit{DB Netz AG} \\
Frankfurt, Cologne and Berlin, Germany \\
Email: \IEEEauthorrefmark{4}{gisela.freiin-von-arnim@deutschebahn.com}, \IEEEauthorrefmark{5}{pedro.lehmann-ibanez@deutschebahn.com},\\ \IEEEauthorrefmark{6}{\{firstname.lastname\}}@deutschebahn.com}
}

\maketitle

\begin{abstract}
The digitalization of railway systems should increase the efficiency of the train operation to achieve future mobility challenges and climate goals.
But this digitalization also comes with several new challenges in providing a secure and reliable train operation.
The work resulting in this paper tackles two major challenges.
First, there is no single university curriculum combining computer science, railway operation, and certification processes.
Second, many railway processes are still manual and without the usage of digital tools and result in static implementations and configurations of the railway infrastructure devices.

This case study occurred as part of the Digital Rail Summer School 2021, a university course combining the three mentioned aspects as cooperation of several German universities with partners from the railway industry.
It passes through all steps from a digital Control-Command and Signalling (CCS) planning in ProSig\,7.3, the transfer, and validation of the planning in the PlanPro data format and toolbox, to the generation of code of an interlocking for the digital CCS planning to contribute to the vision of test automation.

This paper contributes the experiences of the case study and a proof-of-concept of the whole lifecycle for the Digital Testfield of Deutsche Bahn in Scheibenberg.
This proof-of-concept will be continued in ongoing and following projects to fulfill the vision of test automation and automated launching of new devices.
\end{abstract}


\begin{tikzpicture}[remember picture,overlay]
  \node[anchor=south,yshift=10pt] at (current page.south) {\fbox{\parbox{\dimexpr\textwidth-\fboxsep-\fboxrule\relax}{
    \footnotesize \textcopyright 2021 IEEE. Personal use of this
    material is permitted. Permission from IEEE must be obtained for
    all other uses, in any current or future media, including
    reprinting/republishing this material for advertising or
    promotional purposes, creating new collective works, for resale or
    redistribution to servers or lists, or reuse of any copyrighted
    component of this work in other works.
  }}};
\end{tikzpicture}%

\section{Introduction}
The digitalization of railway systems does not only mean introducing new devices and computer-based processes, but digitalization also means new challenges in the area of training and education of all involved persons.
Digitalization is a core topic that applies to all areas and connects them in a new manner.
This also affects security-relevant areas like the railway industry.
For this new challenge, deep knowledge in computer science and domain-specific knowledge about railway operation and certification processes needs to be combined.
No single university curriculum combines all these aspects.

The Digital Rail Summer School 2021\footnote{\url{https://hpi.de/drss}} addresses the challenges of the three mentioned dimensions of the digitalization of the railway in the third iteration.
The Digital Testfield of Deutsche Bahn in Scheibenberg, Saxony offers the opportunity to evaluate and test innovations in praxis.
The planned experimental railway station Scheibenberg is the ideal demonstrator for these tests.
The authors of this paper describe their experiences with a continuous digital planning process as an hour-of-birth of a digital twin as part of the Control-Command and Signalling (CCS) lifecycle and their resulting possibilities for new applications in all phases.

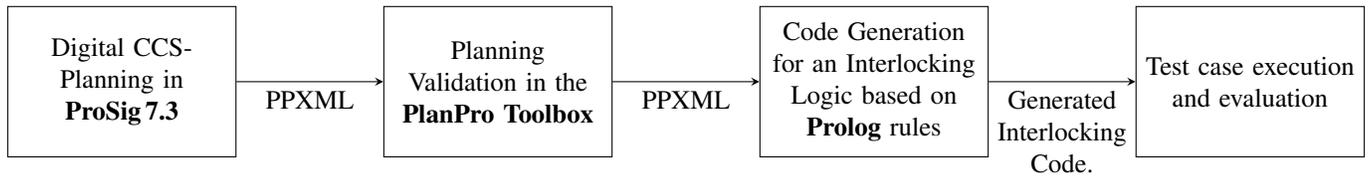
\begin{figure*}
	\centering
	\begin{tikzpicture}
		\node[text centered, draw,rectangle, text width=2.8cm, align=center,minimum height=2cm] (prosig) at (-5,0) {Digital CCS-Planning in \textbf{ProSig\,7.3}};
		\node[text centered, draw,rectangle, text width=2.8cm, align=center,minimum height=2cm] (planpro) at (0,0) {Planning Validation in the \textbf{PlanPro Toolbox}};
		\node[text centered, draw,rectangle, text width=2.8cm, align=center,minimum height=2cm]  (interlocking) at (5,0) {Code Generation for an Interlocking Logic based on \textbf{Prolog} rules};
		\node[text centered, draw,rectangle, text width=2.8cm, align=center,minimum height=2cm]  (testexec) at (10,0) {Test case execution and evaluation};

		\draw[->,-stealth] (prosig) -- node[below]{PPXML} (planpro);
		\draw[->,-stealth] (planpro) -- node[below]{PPXML} (interlocking);
		\draw[->,-stealth] (interlocking) -- node[below, align=center]{Generated\\Interlocking\\Code.} (testexec);

	\end{tikzpicture}
	\caption{The full process of the case study from a digital CCS planning to test automation. The digital CCS planning happens in the tool ProSig\,7.3. This tool produces a PPXML-file following the PlanPro standard containing the planning. In the next step, the PlanPro toolbox validates the planning. This planning is the input for generating the code of an interlocking logic. The vision here is to use such an interlocking to automate the evaluation of new railway infrastructure devices.}
	\label{fig:full-process}
\end{figure*}

This process is visualized in Fig. \ref{fig:full-process}.
Trainees of DB Netze developed, based on digital track network data, an interlocking-planning for the experimental railway station Scheibenberg with ProSig\,7.3 in the PlanPro\footnote{\url{https://www.dbnetze.com/planpro}} data format\cite{maschek2012planpro, klaus_digital_2020}.
This CCS plan is validated in the PlanPro toolbox.
Building on the PlanPro-file, students of the TU Berlin, TU Dresden, and the Hasso Plattner Institute (HPI), Univ. Potsdam created a toolchain to read and process PlanPro files to generate an interlocking logic with Prolog - a common language for artificial intelligence.
This interlocking logic contains first rules like the mutual exclusion of the routes.
This logic will be used as part of the future vision of test automation and automated launching of new devices in the railway domain.

The result of the cooperation of the three universities with DB Netz as part of the Digital Rail Summer School 2021 is a method to generate prototypical interlocking-logics for digital interlockings.
At the same time, this project is an example of the advantages of continuous digital planning processes for the realization and operation of future CCS sites.

The following paper follows the four steps shown in Fig. \ref{fig:full-process}. In the end, it sums up with a conclusion and shows potential for future work.

\section{Status Quo: Digital Planning with PlanPro - Practitioner’s report}
Conventional planning processes for Control-command and Signalling (CCS) technology and even currently widespread "digitalized" solutions, still differ very strongly from digital, data-based CCS planning processes.

\begin{figure*}
	\centering
	\includegraphics{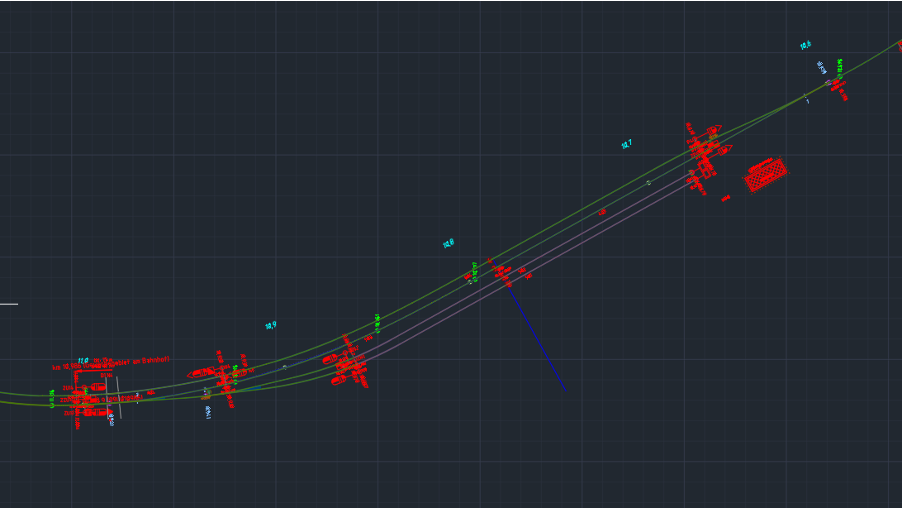}
	\caption{The railway station Scheibenberg of the Digital Testfield of Deutsche Bahn in ProSig\,7.3. The four different tracks together with their signals and endings are visible in the map (best in colors).}
	\label{fig:scheibenberg}
\end{figure*}

Nowadays CCS planning is still strongly based on drawings and paper, with the conventional methods including handwriting and hand-calculations.
The digitalization of these processes has led to the use of self-made excel-sheets as calculation aids as well as PDF documents. Most of the infrastructure engineers today either plan in CAD-based programs such as ProSig\,6 or on paper or proprietary formats of office applications and then send the results to CAD specialists to implement it in ProSig\,6. However, the final products are still PDF documents and hand-filled tables.
To truly achieve a digital planning process, a standardized digital format such as the PlanPro format is needed, so that all the engineering data can be handed over in an XML file and processed automatically. To implement this format, there are several new tools in development. Of the different signal engineering tools, ProSig\,7.3 was used for this particular project, which is shown in Fig. \ref{fig:scheibenberg} with the map of the railway station Scheibenberg. Similar projects will be conducted to test the other available software products. Furthermore, the DB tool PlanPro-Werkzeugkoffer (PlanPro toolbox) was used for visualization, validation, and rule-based quality checking.

To test the PlanPro format as well as the planning tools, a digital CCS planning of the railway station Scheibenberg was conducted as a part of the Digital Testfield of Deutsche Bahn. In the following, the steps taken in this digital planning process will be presented.
The PlanPro-Werkzeugkoffer provided for the current model version 1.9.0.2 was firstly used for the project initialization, which consists of entering the basic project data as well as the provided track layout data. This requires that the latter is available, correct, and converted into the PlanPro format beforehand.
The initialized project, in form of a PlanPro XML file, could then be opened with ProSig\,7.3, in which the actual planning takes place. Since both, the Werkzeugkoffer and ProSig\,7.3, are still in development this import needed to be handled with care, especially regarding the separation of different planning subcategories, in this case, "Geo" (track layout) and "ESTW" (electronic interlocking).
After a successful import in ProSig\,7.3, the actual CCS was planned and the corresponding data was added to the subcategory "ESTW” by the tool. This includes signals, train detection sections, routes, and other physical or logical elements connected to the interlocking respectively defining its functionality. All information must be provided to enable the supplier to build up the related assets and the interlocking software. However, the generic functionality is not covered and has to be taken from the system specifications. Thus, the generic modeling is largely suitable for the different system specifications of different railway infrastructure managers.

\begin{figure*}
	\centering
	\includegraphics[width=0.9\textwidth]{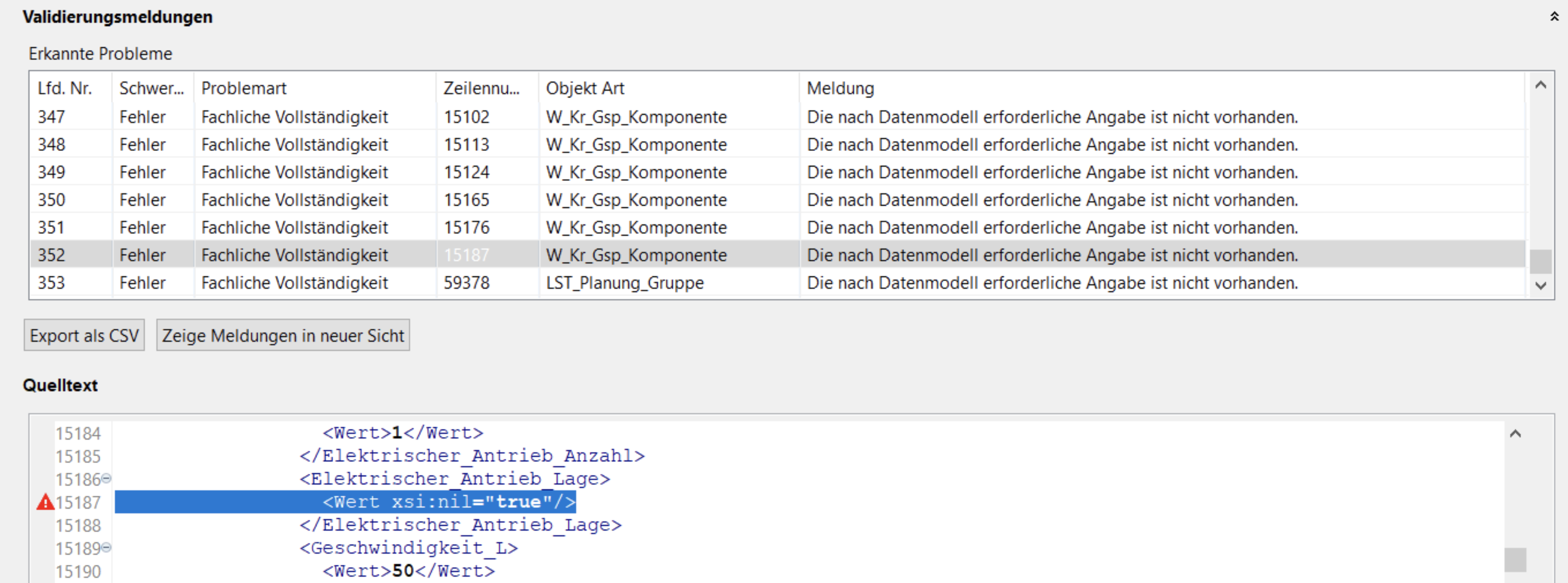}
	\caption{The validation of the Scheibenberg CCS planning in the PlanPro toolbox. During the validation of the Scheibenberg example, serval issues were found in that state of the planning. This toolbox verifies the validity of the planning document.}
	\label{fig:planprovalidation}
\end{figure*}
Then, the finalized engineering data were exported from ProSig\,7.3 as a PPXML file to be imported again into the Werkzeugkoffer for the final schema validation as shown in Fig. \ref{fig:planprovalidation} and a rule-based quality check using Schematron technology\cite{klaus_automated_2018}. The ProSig-internal validation function which shows remaining technical errors as well as schema errors was already used before to close gaps and correct the data in close internal iterations.

Finally, the Werkzeugkoffer creates all relevant engineering tables automatically. The creation of layout and schematic layout plans is still under development so that these documents have to be created from the applied engineering tool.
In the foreseeable future, it will be possible to export all relevant documents for the CCS planning directly from the Werkzeugkoffer.

For now, during a project, close communication with the developers of the digital planning tools is indispensable since there are still many development activities on their end.
Some functions can only be
accessed by the developers and there are still some schema
deviations present in both programs. The span of reasons is
wide: model changes that haven’t been taken over yet,
or random bugs and dynamic reactions which couldn’t be identified
due to a lack of test projects. As an example, many XML-schema errors present in the final version
of this project were due to different numbers of decimal
places for a single property. Furthermore, it is not always clear
which information has to be filled in for a correct and valid
result and which information is irrelevant and might lead to
errors. In this case, the translation of the generic language
and variability of the model into meaningful input forms and
processes is the challenge. It can only be met when IT and
signaling competence are put together and adequate testing
is granted. Therefore, this project gave important input and
served as a feedback loop to the developers, to further refine
their digital planning software. Due to the very manageable
size of the station, the complete engineering process could be
gone through in a short time. This way, fundamental issues
related to interfaces and process steps could be identified and
handled quickly without being confronted with an overloading
amount of error occurrence.
Therefore, this project gave important input and served as a feedback loop to the developers, to further refine their digital planning software. Due to the very manageable size of the station, the complete engineering process could be gone through in a short time. This way, fundamental issues related to interfaces and process steps could be identified and handled quickly without being confronted with an overloading amount of error occurrence.

If the previously described issues were to be solved and more importantly, there was complete and reliable data concerning track topology and existing CCS elements of today’s infrastructure, soon digital planning could follow the described workflow and be much faster and far less prone to errors. It would also allow the planning engineers to focus more on the challenging aspects of planning, since the software will help minimize tedious tasks, like hand-filling tables, measuring distances, and many more.

\section{The PlanPro format and its eco-system – the infrastructure manager’s perspective}

The basic approach of DB Netz regarding digital CCS engineering is to define the data format for the interfaces along the engineering process based on the company’s uniform object model. Although a common, company-wide accepted and “complete” model could not be achieved so far, PlanPro delivers important input for the CCS content and has proven its stability in many activities.
Although PlanPro doesn’t include 3D representations it is fully in line with the BIM approach, which focuses on consistent and continuous data management as a basis for high-quality results and a transparent engineering process. By that, PlanPro builds the fundament for data-driven business models in the CCS context which will dominate in the future.
The binding provision of the interface goes along with the possibility of opening the market for different engineering tools. Although several activities can be seen now, the provision of suitable tools for the whole engineering chain is still a challenge. Additionally, the migration of projects and the digitalization of the inventory have to be solved to achieve the full advantages of digital work. It has to be decided soon, if a central digitalization process can be the solution or if data will be created project by project. The latter has been chosen so far with the disadvantage of long migration paths and time.
To pave the way for digital transformation the engineering rules are being adapted stepwise. The first steps were made in 2017 with technical instructions for operational testing \cite{morgenroth_tm_2017}. The results have recently been integrated into the engineering rules which now describe the conventional and the digital way in parallel. The publication is planned for later this year.
Besides these technical and processual regulations, a certain challenge was and still is the acceptance of the new methodology. Therefore, legal requirements and official regulations must develop in parallel, same as contractual conditions and the attitude to innovation.

\section{From the validated digital planning to code generation for an interlocking logic - an IT-system engineers perspective}

The next step in the process in Fig. \ref{fig:full-process} is the generation of code for an interlocking logic.
This interlocking logic is a core part of a test automation suite and should manage a railway section.
This interlocking logic bases on the previously created and validated CCS planning of the experimental railway station Scheibenberg in the standardized PlanPro format.

\begin{figure*}
	\centering
	\begin{tikzpicture}
		\node[text centered, draw,rectangle, text width=2.8cm, align=center,minimum height=2cm]  (interlocking) at (0,0) {Generic Interlocking based on \textbf{Prolog} rules};
		\node (planpro) at (-4,0) {};
		\node (testexec) at (4,0) {};

		\draw[->,-stealth] (planpro) -- node[below]{PPXML} (interlocking);
		\draw[->,-stealth] (interlocking) -- node[below, align=center]{Interlocking\\Logic} (testexec);

		\node[text centered, draw,rectangle, text width=2.8cm, align=center,minimum height=1.5cm]  (parse) at (-5,-3) {Parse PlanPro PPXML file};
		\node[text centered, draw,rectangle, text width=2.8cm, align=center,minimum height=1.5cm]  (generate) at (0,-3) {Generate all Routes};
		\node[text centered, draw,rectangle, text width=2.8cm, align=center,minimum height=1.5cm]  (interlocking-rules) at (5,-3) {Code~Generation for Interlocking Logic};
		\node (input) at (-8.5,-3){};
		\node (output) at (8.5,-3){};

		\draw[->,-stealth] (input) -- node[below]{PPXML} (parse);
		\draw[->,-stealth] (parse) -- node[below,align=center]{Prolog\\Facts} (generate);
		\draw[->,-stealth] (generate) -- node[below,align=center]{Facts \&\\ Routes} (interlocking-rules);
		\draw[->,-stealth] (interlocking-rules) -- node[below,align=center]{Generated\\Interlocking\\Code} (output);

		\draw[dotted, transform canvas={xshift=-0.2cm, yshift=0.2cm}, thick, gray!80] ($(interlocking.south west)$) -- ($(input.north east)$);
		\draw[dotted, transform canvas={xshift=0.2cm, yshift=0.2cm}, thick, gray!80] ($(interlocking.south east)$) -- ($(output.north west)$);

	\end{tikzpicture}
	\caption{An overview of the EULYNX-Live project. The PlanPro parser reads a CCS planning document. Afterward, a Prolog program creates all routes in the railway network and in the last step, code for an interlocking is generated. This is the basis for the test execution.}
	\label{fig:eulynx-live}
\end{figure*}
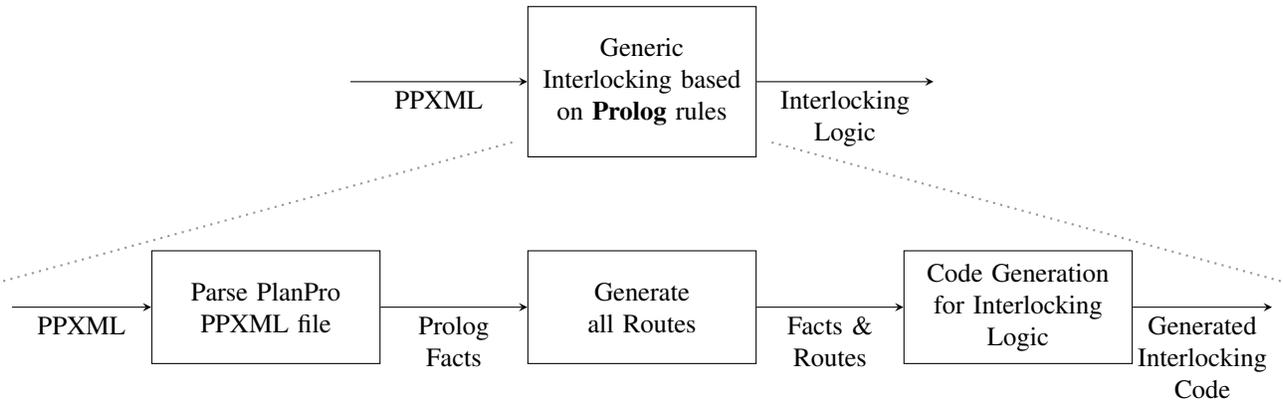

The process to create the interlocking logic based on the PPXML file is displayed in Fig. \ref{fig:eulynx-live}.
We decided to use Prolog for the whole process since Prolog is a common language for artificial intelligence.

The first step is the parsing of the PlanPro file.
In this step, all facts about the railway section are added to the Prolog database.
This includes the topology of the railway network represented as a graph of nodes and edges.
Every node represents a point or a track-end; every edge is a track.
These edges also contain the detail, how they are connected to the point.
Besides the topology, also the infrastructure elements like signals are part of the PPXML file.
The signals are connected to an edge in a specific direction to the edge.

Based on these facts, we are generating all routes through the railway network.
To generate the routes, every start signal is a potential start point and a set of \textit{interesting points}, mostly end signals, are potential endpoints of a route.
Starting from the start points, the topology graph is traversed in the next step.
Every time the algorithm is reaches an endpoint, a new route is created.
This set of routes is a superset of all specified routes and operating procedures in that railway network.
We are creating this superset to ensure, that all possible (future) combinations are handled and validate, that the CCS planner hasn't forgotten any route.
This is another validation capability of a digital planning process.
Each route will be a test case in the test automation.

Afterward, rules for the mutual exclusion of the routes were implemented.
Through running these rules, code for a first simple interlocking logic based on a state machine can be generated.
Additional rules are easy to add here.
This interlocking logic is a digital twin of a real interlocking and as part of the test automation, this logic will manage a railway section.

\section{From an interlocking logic to test automation - a test engineers perspective}

The big picture of this case study is to automate the creation and execution of tests in the railway infrastructure domain.
Due to new open standards, like from the EULYNX consortium, many more manufacturers are entering the market with new products.
An automated way to test the interoperability with other products from other manufacturers is a key challenge to get new products to production.

\begin{figure}
	\centering
	\begin{tikzpicture}
		\node[draw, align=center, minimum height=1.5cm, minimum width=2.2cm] (orchestrator) at (0,0.5) {Lab Orchestrator};
		\node[draw, align=center, minimum height=1.5cm, minimum width=2.2cm] (hpi) at (-3,-2.5) {HPI IoT-Lab\\Potsdam};
		\node[draw, align=center, minimum height=1.5cm, minimum width=2.2cm] (ebuef) at (-1.5,-4.5) {EBueF\\TU Berlin};
		\node[draw, align=center, minimum height=1.5cm, minimum width=2.2cm] (btc) at (0,-2.5) {BTC\\Wustermark};
		\node[draw, align=center, minimum height=1.5cm, minimum width=2.2cm] (brunel) at (1.5,-4.5) {Brunel\\Hildesheim};
		\node[draw, align=center, minimum height=1.5cm, minimum width=2.2cm] (scheibenberg) at (3,-2.5) {Digital\\Testfield\\Scheibenberg};

		\node[draw, align=center, minimum height=0.5, minimum width=8.2cm] (internet) at (0,-1) {Internet};

		\draw (internet) -- (orchestrator);
		\draw (-3,-1.25) -- (hpi);
		\draw (-1.5,-1.25) -- (ebuef);
		\draw (0,-1.25) -- (btc);
		\draw (1.5,-1.25) -- (brunel);
		\draw (3,-1.25) -- (scheibenberg);

	\end{tikzpicture}
	\caption{The overview about the \textit{Distributed IoT-Lab}. Besides the central orchestration node, it shows the different sides of the laboratory, which are connected over the Internet. This enables the capability to connect object controllers from different physical-separated locations in a single test case.}
	\label{fig:distributed-lab}
\end{figure}
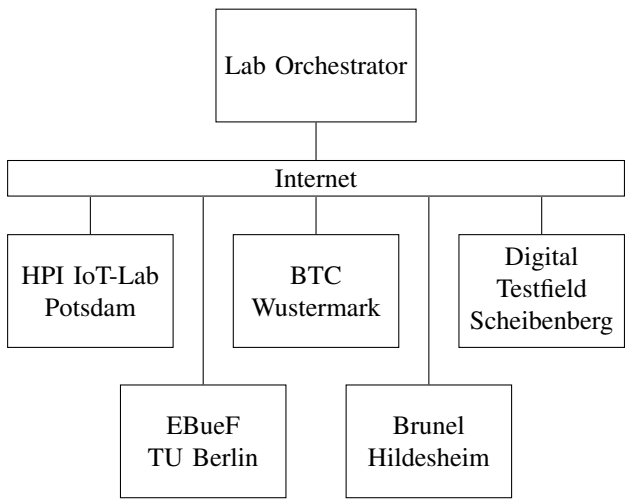

Each route in the railway network defines a test case.
The idea is to use such a test case as the scenario and use devices and implementations of different manufacturers as operating parts of that scenario.
Each operating part, like an object controller of an infrastructure element or an interlocking implementation, can be either virtual or in hardware.
Afterward, the railway network acts like described in the test case. A monitoring instance evaluates the behavior of the devices. After the execution, the performance of the new device can be evaluated.

To have flexible and realistic test executions, the EULYNX-Live concept \cite{schmid2021eulynx}, a triad of simulation, testbed, and field test is used.

In the early stages of development, the software of the new device will be executed in a simulation-based environment. This means every operating path is virtualized with a digital twin, which leads to flexible and scalable test executions due to the usage of simulation. A huge number of different scenarios can be executed in parallel. This can easily be done by automation since the software can easily be reconfigured.

A disadvantage of simulation is the applicability to reality. Each simulation has some abstractions of reality. The biggest abstraction for the test execution is the not-existence of real hardware devices. This leads to missing hardware effects. A way to solve this is to use hybrid testbeds like Marvis \cite{beilharz2021towards} for the test execution. A hybrid testbed offers the capability to use hardware-in-the-loop techniques to include the device-under-test into a simulated test environment. This leads to more realistic test cases, but reduces the flexibility, especially prohibit the parallel execution of test cases. In a hybrid testbed scenario, only the device-under-test is in hardware. Every other operating part can be virtualized to keep flexibility on that end and have scalable test cases. To increase the realism of the test execution additional hardware devices can be integrated. This relates to the challenge to have a large set of devices.
To tackle this challenge, we created the \textit{Distributed IoT-Lab} as shown in Fig. \ref{fig:distributed-lab}.
This distributed laboratory connects several laboratories at other research institutes and infrastructure manufactures.
A central lab orchestrator manages all connected devices.
This connection allows us to include several real hardware devices in our test environment.
In combination with the EBuEF\footnote{\url{https://www.ebuef.de/}}, complex scenarios can be tested \cite{herholz_echter_2020}.

But even in a hybrid testbed, not all hardware devices are fully integrated. Hardware effects only affect the device under test.
Finally, a field test is still necessary.
In addition to tests in a hybrid testbed, real trains can be used to check real scenarios completely.
Currently, the Digital Testfield of Deutsche Bahn Scheibenberg is the only conceivable location for this due to the capabilities of a non-productive but modern equipped testing area.
In this way, real results are obtained and errors are found that cannot be detected in the hybrid test field, due to not all hardware components were involved.
The disadvantage is that not all scenarios can be tested. Test cases that produce exceptions can cause damage. In addition, the tests and also the errors in the field test are more expensive, e.g. by the provision of the trains.
It is also difficult to automate the tests. Nowadays, the commissioning of CCS systems takes place manually. Nevertheless, approaches can be found here, such as evaluating signal aspects with image processing. Signals can be detected with
a drone and then the shown signal aspect can be evaluated and documented using image processing algorithms.
With this combination of simulation, hybrid testbed, and field test, a lean test-driven development process can be implemented.

\section{Conclusion and Future Work}

To put digital Control-Command and Signalling (CCS) systems at work, expertise from multiple domains has to be combined. Future CCS engineers have to combine knowledge in computer science with expertise in railway operations and certification. In contrast to traditional safety-certified systems, that will be left untouched for long periods (several years), digital systems have to address the problems of IT security using appropriate encryption techniques. However, to stay secure, IT systems will need to be updated continuously. Future certification processes will need to take this into account.

Dependencies on tools and tool manufacturers describe another problem that arises when building predictable, reliable, and safe IT systems. Reliability assessments of future systems will need to evaluate the predictability and trustworthiness of toolchain manufactures. Formal methods and code generation appear as promising approaches.

The digital planning process offers the capabilities for CCS planners to have an early-as-possible validation of planning documents due to several layers of validation with PlanPro and EULYNX-Live.
Additionally, infrastructure providers and manufacturers also benefit from the new possibilities for interoperability and integration tests.

Within this paper, we discussed how fully digitalized processes covering the entire lifespan of field elements can be implemented. Part of the digital planning process is the creation of digital twins of our environments. A digital twin for existing field elements will give us the chance, to implement automated checking and predictive maintenance approaches – that have great potential for future systems. At the same moment, one has to be aware of new risks of vendor lock-in and dependencies on manufacturers of digital planning tools.

\section*{Acknowledgment}

We would like to thank our colleagues, students, and friends from TU Berlin, TU Dresden, HPI, and others who contribute to Digital Rail Summer School 2021.
Especially we would like to mention Raik Hoffmann in his role as coordinator of the Digital Testfield of Deutsche Bahn in Scheibenberg.
We acknowledge the continuing support by DB Netze I.NAT3 - namely Dr. Bernd Elsweiler, Wolfgang Klein, and Jens Reinel.
Part of this work has been funded by the German Federal Ministry of Transport and Digital Infrastructure through the mFUND project RailChain (19F2093C).

\bibliographystyle{unsrt}
\bibliography{refs}


\end{document}